\documentclass{webofc}
\usepackage[varg]{txfonts}   
\begin{document}
\title{Nuclear clustering and the electron screening puzzle}
\author{\firstname{C.A. } \lastname{Bertulani}\inst{1}\fnsep\thanks{\email{carlos.bertulani@tamuc.edu}} 
\and
\firstname{C. } \lastname{Spitaleri}\inst{2,3}\fnsep\thanks{\email{spitaleri@lns.infn.it}}
}
\institute{
Department of Physics and Astronomy, Texas A\&M University-Commerce, Commerce, Texas 75429, USA  \and
Department of Physics and Astronomy, University of Catania, Catania, Italy \and
INFN-Laboratori Nazionali del Sud, Catania, Italy 
}

\abstract{%
Electron screening changes appreciably the magnitude of astrophysical nuclear reactions within stars. This effect is also observed in laboratory experiments on Earth, where atomic electrons are present in the nuclear targets.  Theoretical models were developed over the past 30 years and experimental measurements have been carried out to study electron screening in thermonuclear reactions. None of the theoretical models were able to  explain the  high values of the experimentally determined screening potentials. We explore the possibility that  the ``electron screening puzzle" is due to nuclear clusterization and polarization effects in the fusion reactions. We will discuss the supporting arguments for this scenario. 
}
\maketitle
\section{Introduction}
\label{intro}
Thermonuclear reactions in stars occur at ultra-low energies  around the Gamow energy, E$_G$ , of the order of few keV to 100 keV, much below the Coulomb barrier \cite{Rolfs1988,Adelberger2011}. The experimental determination of the fusion cross sections for these reactions, $\sigma_b (E)$, requires the use of either direct or indirect techniques \cite{BG10,TBC14}.  In the Sun, most nuclear reactions involve charged  particles and non-resonant reactions \cite{Adelberger2011,BK16}.
As the relative energy between ``bare" nuclei decrease, their cross sections $\sigma_b$(E) decrease exponentially and it is more appropriate to express the cross section in terms of  the astrophysical S-factor, defined as
\begin{equation}
 \sigma_b(E)={1\over E}  S(E) \exp\left[-2\pi\eta(E)\right] , \label{sbe}
\end{equation}
where $1/E \propto \pi \lambdabar^2$ accounts for the quantum mechanical area $\pi \lambdabar^2$ within which the reaction can occur, with $\lambdabar$ being the reduced wavelength and  $\exp(-2\pi\eta)$ is a function which approximately accounts for the tunneling probability due through the barrier.
The  Sommerfeld parameter $\eta(E)$   is given by
\begin{equation}
\eta(E) = \frac{Z_1Z_2e^2}{\hbar v}=\frac{Z_1Z_2  \alpha}{v/c} ,  \label{eta}
\end{equation}
where Z$_1$, Z$_2$ are the nuclear charges, $v=\sqrt{2E/\mu}$ the initial relative velocity  between the nuclei, and $\alpha=e^2/\hbar c$ is the fine-structure constant. With this definition, the astrophysical S-factor becomes a much smoother function of $E$ than $ \sigma_b(E)$ allowing for a more reliable extrapolation from the high energy values, where measurements (direct or indirect) have been carried out, to  the low energies around Gamow peak \cite{Adelberger2011}.

The presence of electrons in atoms increases the tunneling probability, also increasing the measured (screened) cross-section $\sigma_s$. Free electrons in stars also induce and increase of the cross sections  \cite{Assenbaum1987,Rolfs1988,Fiorentini1995,Strieder2001}.
One can define an enhancement factor $f_{lab}(E)$ in laboratory experiments by means of \cite{Rolfs1988,Fiorentini1995, Strieder2001},
\begin{eqnarray}
f_{lab}(E) = \frac{ \sigma_s(E)}{\sigma_b(E)} = \frac{S_s(E)}{S_b(E)}  \simeq \exp\left[\pi\eta\frac{U_e{^{(lab)}}}{E}\right], \label{fe}
\end{eqnarray}
where $U_e^{(lab)}$ is the electron screening potential.
This enhancement factor has been studied experimentally, but it was found out that in many cases the measured values extracted for $f_{lab}(E_G)$ are in significant disagreement with atomic-physics models  \cite{Rolfs1988, Strieder2001}.       
In this proceedings contribution we report a  novel approach  to explain this puzzle with nuclear  clustering  effects, as recently shown in Ref.  \cite{Spit16}. 

\begin{figure}
\begin{center}
\includegraphics[scale=0.28]{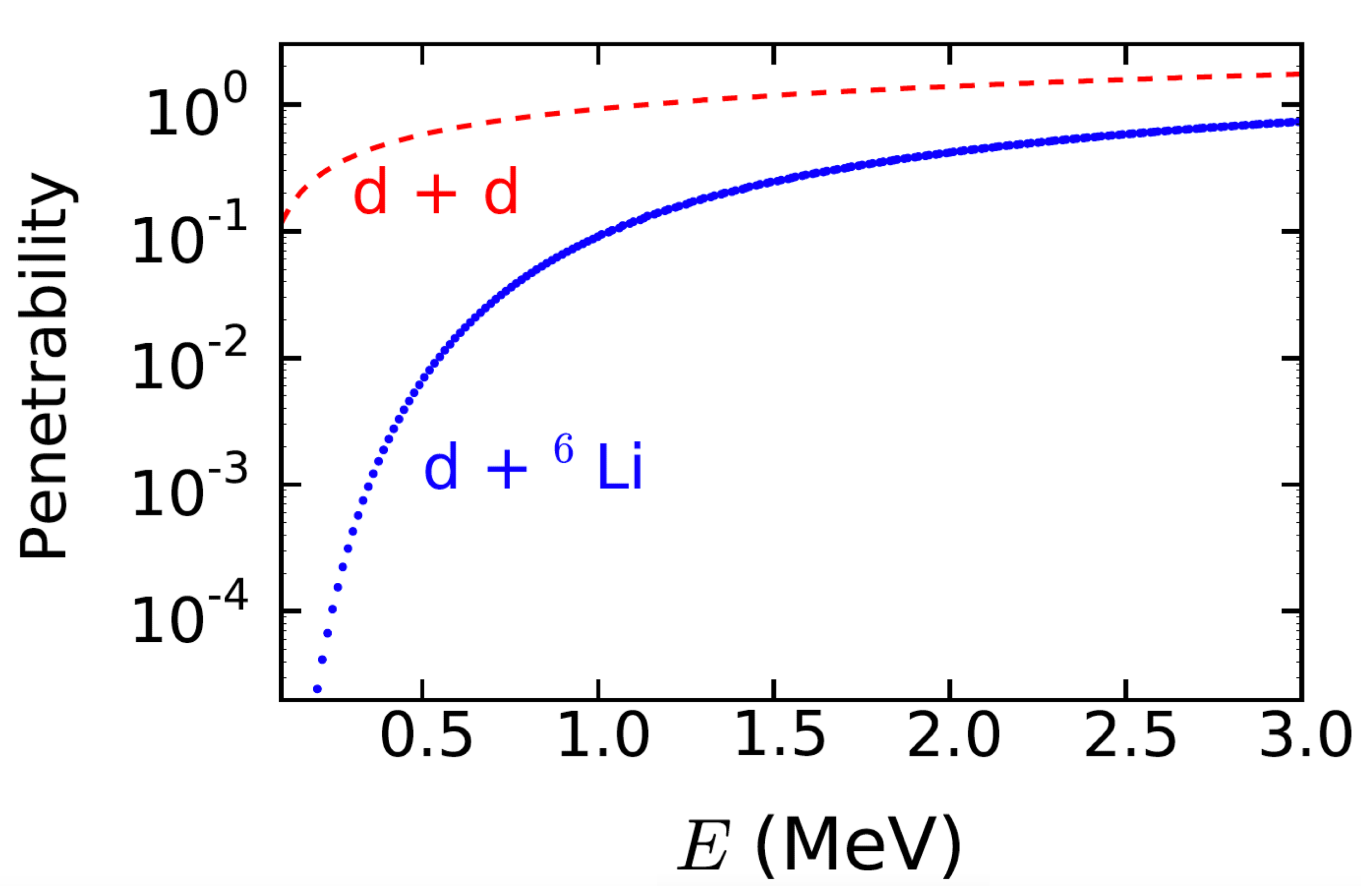}\label{fig1}
\caption{Barrier penetrabilities for d + d and for d + $^6$Li reactions as a function of the relative motion energy.}
\end{center}
\end{figure}

\begin{figure}
\begin{center}
\includegraphics[scale=0.22]{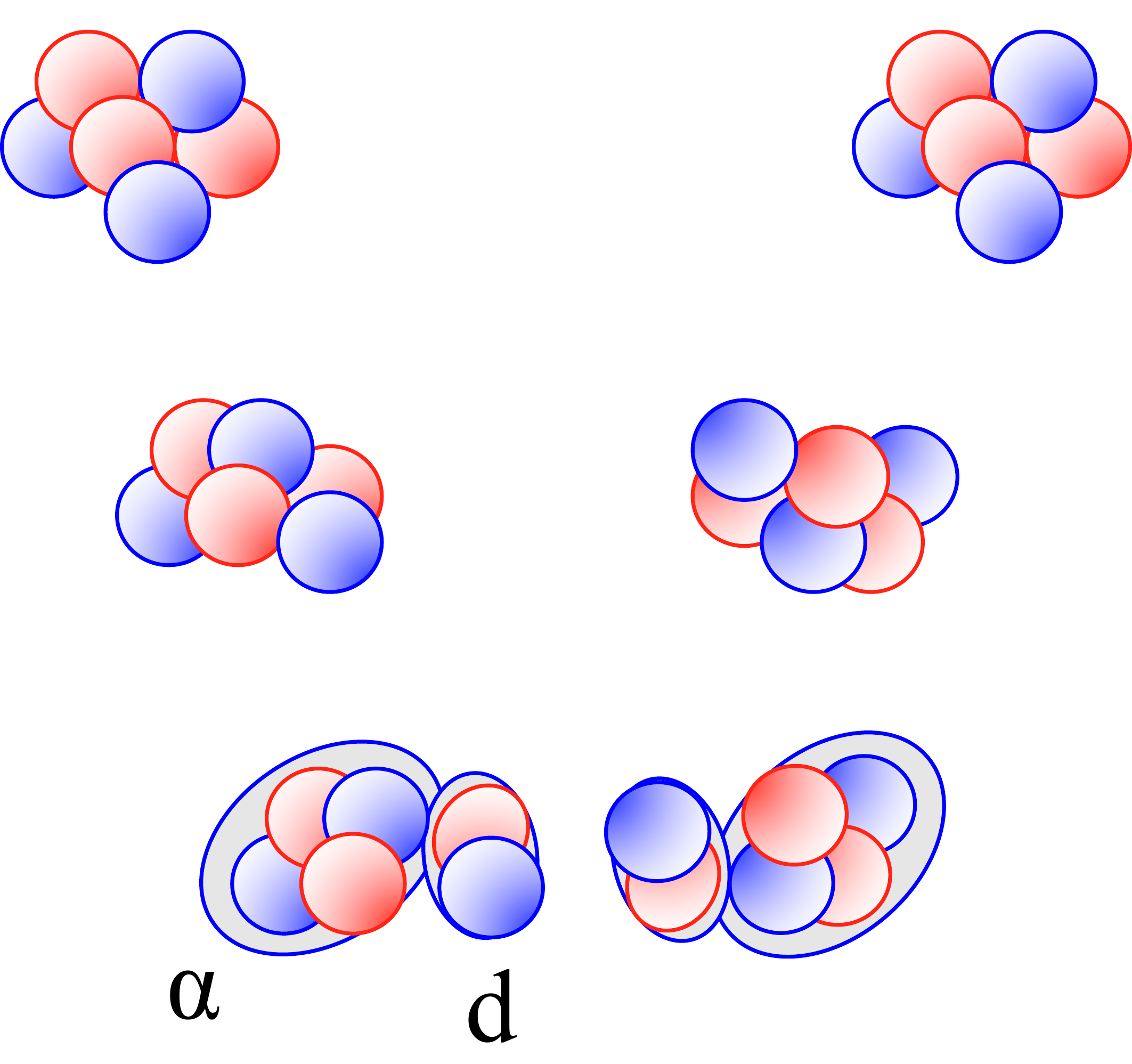}
\end{center}
\caption{Schematic view of polarization and orientation as nuclei with large probabilities for cluster-like structures approach each other, shown here  for $^6$Li + $^6$Li.}\label{fig1b}
\end{figure}

In the Fock space, a properly anti-symmetrized nuclear wavefunction can be written as
\begin{equation}
\left |\psi_{nucleus}\right> = \alpha\left|\psi_A\right> + \beta\left|\psi_a\psi_B\right> + \gamma\left|\psi_c\psi_D\right> + \cdots ,\label{wf} 
\end{equation}
where  $\left|\psi_A \right>$ is a single-cluster wave function of  $A$ nucleons,  $\left|\psi_a\psi_B\right>$ is a cluster (a+B) nuclear wave function  and ($\alpha, \beta, \gamma$) are spectroscopic amplitudes for such configurations.  Since the Coulomb penetrability changes for each cluster configuration, fusion probabilities are effectively reduced, increasing the cross sections and leading to a similar effect as that of electron screening. We illustrate this with one example.  Experiments show that the  $^6$Li + $^6$Li $\to$ 3$\alpha$ cross section is  orders of magnitude larger than expected from barrier penetrabilities  based on a non-clusterized $^6$Li nucleus \cite{Lattuada1988}.
But  $^6$Li has a non-negligible probability to be found as $d + \alpha$ cluster structure. With this organized structure, fusion is enhanced as the Coulomb barrier for deuterons with $^6$Li is reduced.
The cross section can be written as
\begin{equation}
\sigma_L = C_{66}P_{^6{\rm Li}+^6{\rm Li}} + C_{26} P_{{\rm d}+^6{\rm Li}} + C_{46} P_{^4{\rm He}+^6{\rm Li}}+ \cdots , \label{slc}
\end{equation}
where $C_{66}$,  $C_{26}$, etc, are variables that include the energy dependence not included in the tunneling probabilities for the  $^6$Li + $^6$Li,  d + $^6$Li, etc, configurations. Certainly, ${P_{^6{\rm Li}+^6{\rm Li}}}/{P_{{\rm d}+^6{\rm Li}}} \to 0   $ as $k \to 0$, and the second term in Eq. (\ref{slc}) will be dominant as $E \to 0$. The energy dependence of these variables are much weaker than of the barrier penetrabilities and therefore  cluster-like structures will enhance the cross sections by many orders of magnitude. Moreover, if the deuterons within the $^6$Li nuclei approach each other and directly fuse to $\alpha$ particles, it would increase enormously the cross section for the  reaction,  as shown in Figure  \ref{fig1}. That is what one observes with the cross section for $^6$Li + $^6$Li $\to$ 3$\alpha$ \cite{Manesse1964,Gadeken1972,Lattuada1987,Lattuada1988,Spitaleri2015}.  It is likely that the deuterons within $^6$Li tunnel through the barrier individually to form the $\alpha$-particles, leaving the other two $\alpha$-particles behind.  
The enhancement is probably helped by polarization, because $^6$Li nuclei align so that the two deuterons can be located  closer to each other during the  tunneling process, thus reducing the effective barrier. The alpha particles stay farther away from each other due to polarization and alignment, as schematically shown in Figure \ref{fig1b}. An average over all configurations leads to a reduced effective Coulomb repulsion. This was demonstrated using a simple cluster  model for $^6$Li in Ref. \cite{Spit16}.

\begin{figure}
\begin{center}
\includegraphics[scale=0.38]{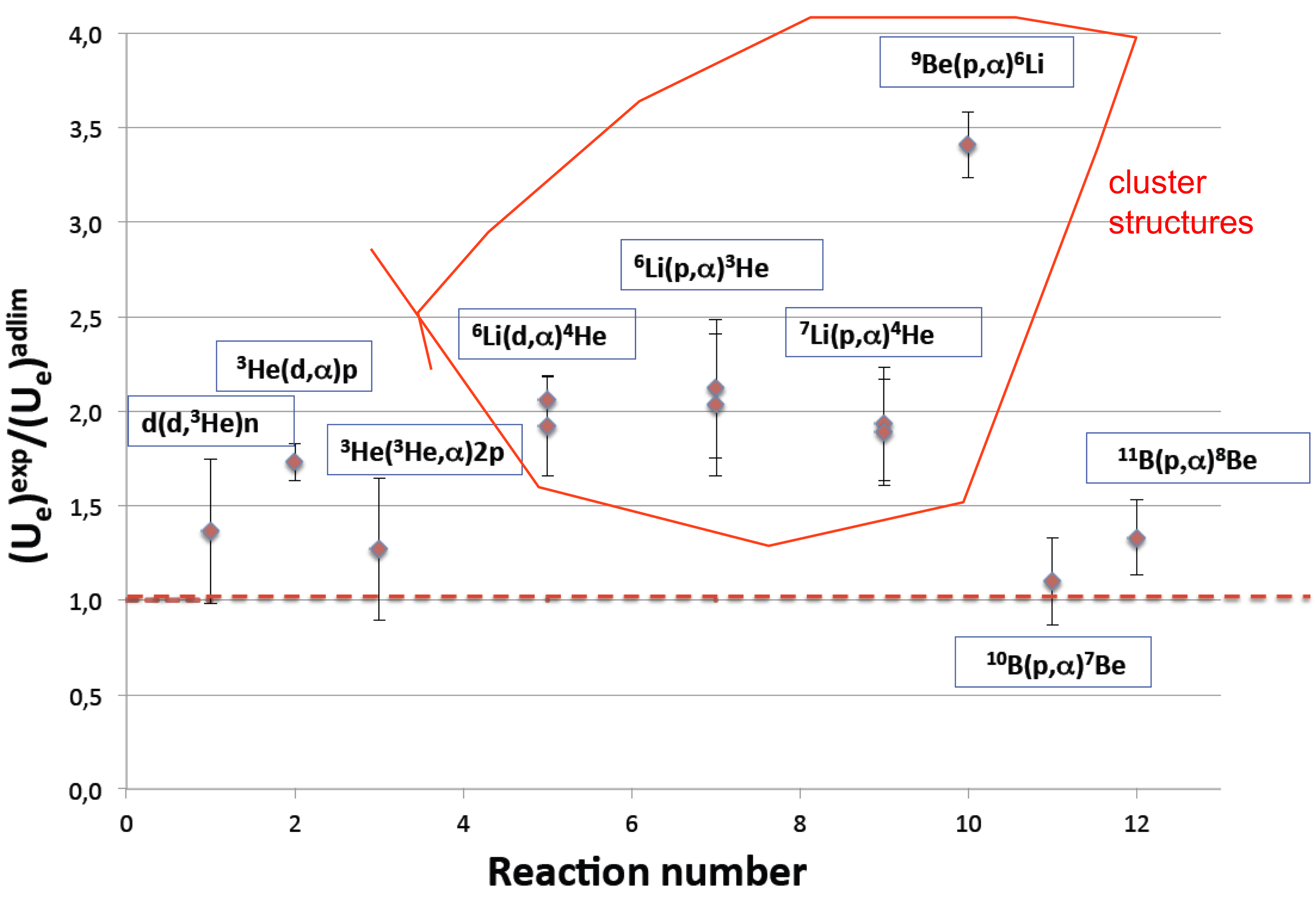}
\end{center}
\caption{Ratio of experimentally determined electron screening potential $U_e^ {exp}$  and the adiabatic limit $U_e^ {a dlim}$  for the reactions listed in Table 1. The reactions involving cluster-like nuclei are encircled by a sketchy line.} 
\label{fig2}
\end{figure}

The model described in Ref. \cite{Spit16} clearly proves that the action of both polarization and orientation, always enhances fusion in the dicluster-dicluster configuration compared to the  non-cluster configuration. Polarization is remnant of the  Oppenheimer-Phillips effect \cite{OpPh,Greife}, which has been used before to quantify  deuteron polarization effects in the Coulomb field of a nucleus in reactions induced by deuterons. But little can be said about the amplitudes $C_i$ in Eq. \eqref{wf}. There is no robust theory with a predictive power for the clustering amplitudes in nuclei, despite intensive theoretical efforts beyond the na\"ive shell-model \cite{Free17}. The nuclear clustering theory also needs to account for cluster pre-formation amplitudes, which needs to be tackled with in ab-initio models (see, e.g., Ref. \cite{NavQua,SIC13,Xu17}).  But even without information about pre-formation factors, or spectroscopic amplitudes for clustering configurations, the mere assumption of the existence of clusters in nuclei will lead to an effective enhanced tunneling when averaged over all orientations \cite{Spit16}. 

\begin{table}
\caption{Experimentally  determined electron screening potentials, $U_e^{exp}$,
 and  the adiabatic limits, $U_e$$^{adlim}$.}
\begin{center}
\begin{tabular}{lllccr}
\hline
{}&                                                                                             {}&                                               {}&                                     {}&            					          {}&                        \\[0.ex]	
{}&Reaction      				       	  	                                 {}&$U_e^{adlim}$     	             {}&$U_e^{exp}$               {}&Note    		     				  {}&Ref.                  \\[0.ex]  
{}&                  					                                     	{}& (eV)                 	             {}&(eV)                             {}&	                         		         {}&                         \\[0.ex]           
\hline
$[1]$  {}&$^2$H($d,t$)$^1$H                                                      {}&14       {}&19.1$\pm$3.4       {}&                                                  						   {}&\cite{Greife,Tumino2014}           \\[0.ex]       
$[2]$ {}&$^3$He($d$,$p$)$^4$He                                              {}&65        {}&109$\pm$9           {}&D$_2$ gas target          		  						   {}&\cite{Aliotta2001}    				    \\[0.ex]   
$[3]$ {}&$^3$He($d$,$p$)$^4$He                                              {}&120      {}&219$\pm$7            {}&        	         	                                         			          {}&\cite{Aliotta2001}    	 			   \\[0.ex]       
$[4]$ {}&$^3$He($^3$He,2p)$^4$He                                          {}&240     {}&305$\pm$90          {}& compilation    		                   						    {}&\cite{Adelberger2011} 			    \\[0.ex]
\hline
$[5]$  {}&$^6$Li($d$,$\alpha$)$^4$He                  		      	       {}&175     {}&330$\pm$120         {}&H gas  target      				                    				 {}&\cite{Engstler1992}    \\[0.ex]
$[6]$  {}&$^6$Li($d$,$\alpha$)$^4$He                  		              {}&175     {}&330$\pm$49           {}&	         		                                                                        {}&\cite{Engstler1992, Musumarra2001}          \\[0.ex]
$[7]$ {}&$^6$Li($p$,$\alpha$)$^3$He                 		               {}&175     {}&440$\pm$150         {}&H gas target       			 		        			        {}&\cite{Engstler1992}          \\[0.ex]
$[8]$ {}&$^6$Li($p$,$\alpha$)$^3$He                   	                       {}&175     {}&355$\pm$67         {}&        													{}&\cite{Engstler1992,Cruz,Lamia2013}           \\[0.ex]                      
$[9]$ {}&$^7$Li($p$,$\alpha$)$^4$He                     		        {}&175     {}&300$\pm$160          {}&H gas target       			 	             			               {}&\cite{Engstler1992}           \\[0.ex] 
$[10]$ {}&$^7$Li($p$,$\alpha$)$^4$He                      		        {}&175     {}&363$\pm$52          {}&      			 	                  						        {}&\cite{Engstler1992,Cruz,Lamia2012a}           \\[0.ex]   
\hline				
$[11]$  {}&$^9$Be($p$,$\alpha_0$)$^6$Li                                 {}&240       {}&788$\pm$70             		{}&  	                                        								{}&\cite{Zahnow1997,Wen2008}             \\[0.ex]                                                                                   
                                          
$[12]$  {}&$^{10}$B($p$,$\alpha_0$)$^7$                                {}&340      	{}&376$\pm$75           {}&                           										     {}& \cite{Angulo1993,Spitaleri2014}     	       \\[0.ex] 
$[13]$  {}&$^{11}$B($p$,$\alpha_0$)$^8$Be        		             {}&340     {}&447$\pm$67            	  {}&            	          	     										 {}& \cite{Angulo1993,Lamia2012}             \\[0.ex]  
\hline	
\end{tabular}
\end{center}
\end{table}

Typical cases of experimentally studied thermonuclear reactions are shown in Table 1  and Figure \ref{fig2}. We claim that in most of these cases, clusterization fusion enhancements have already been observed. Our argument is based on the fact that a clear separation exists for $Z=1$ nuclei reactions with nuclei without cluster sub-structures, and  those with  cluster-like  nuclei. There  exists a discrepancy between the adiabatic limit of the screening potential, $U_e^{adlim}$, and  its experimental value, $U_e$$^{exp}$, but the disagreement is larger when a cluster structure is expected.
The values displayed in  Table 1 favor a nuclear clustering solution for the electron screening problem at the  thermonuclear energies of astrophysical interest. To summarize, the evidences are: (a) Reactions involving non-cluster-like nuclei with a ($Z=1$) nucleus display an agreement, within experimental errors,  with the adiabatic theory of the screening energy \cite{Assenbaum1987} (e.g.,  the d + d and d + p reactions), (b) Reaction cross sections of cluster-like nuclei  with $Z=3$ and $A=6,7$ (e.g.,  p + $^{6,7}$Li)  are more than a factor 1.5 larger than expected using the theoretical U$_e$$^{adlim}$, (c) Experimental values of $U_e$$^{exp}$ for reactions with $Z=4$ cluster nuclei clearly disagree with the adiabatic approximation. And this discrepancy increases with nuclei with more evidenced cluster structures (e.g.,  p + $^9$Be and p + $^{10,11}$B).

We conclude that  the electronic screening problem might be due to a nuclear structure effect in the form of polarization and orientation of nuclei during the fusion process and we propose that new studies be carried out to confirm our claims. Candidates for such a search can involve reactions with $^{6}$Li or $^{7}$Li. For example, one could envisage studies  with  $^6$Li + $^6$Li,  $^7$Li + $^7$Li,  $^7$Li + $^6$Li , $^9$Be + $^3$He, $^9$Be + $^7$Li at thermonuclear energies. For more details and explicit description of calculations mentioned in this proceedings contribution, see Ref. \cite{Spit16}

\bigskip

\begin{acknowledgement}
We have benefited from useful discussions and a collaboration with A. Vitturi and L. Fortunato. This work has been partially supported by the Italian Ministry of University MIUR under the grant ``LNS-Astrofisica Nucleare (fondi premiali)" and the  U.S. NSF Grant No. 1415656, and U.S. DOE grant No. DE-FG02-08ER41533.
\end{acknowledgement}

\end{document}